\documentstyle[12pt,epsfig]{article}

\newcommand{\eps}{\epsilon}

\newcommand{\beq}{\begin{equation}}
\newcommand{\eeq}{\end{equation}}

\newcommand{\ba}{\begin{array}{c}}
\newcommand{\ea}{\end{array}}
\newcommand{\dis}{\displaystyle}
\newcommand{\bd}{\begin{displaymath}}
\newcommand{\ed}{\end{displaymath}}
\newcommand{\br}{\beq\renewcommand{\arraystretch}{1.3}\begin{array}{l}}
\newcommand{\er}{\ea \renewcommand{\arraystretch}{1}\eeq}

\newcommand{\lab}{\protect\label}

\begin{document}

\begin{titlepage}
\begin{flushright}
BUTP-95/7\\
NORDITA 95/41 N,P\\
hep-ph/9505348\\
\end{flushright}
\vspace{0.5cm} \begin{center}

{ \LARGE \bf Mass of scalar resonances\\ \vspace*{0.5cm}
beyond the large-$N_c$
limit\footnotemark
\footnotetext{Work supported in part by the EEC Human Capital and Mobility
Program}}

\vspace*{1.5cm}
      {\bf C. Bruno\footnotemark
      \footnotetext{ email: cbruno@butp.unibe.ch}}\footnotemark
\footnotetext{also at: INFN-Laboratori Nazionali di Frascati P.O. Box 13,
I-00044 Frascati, Italy}
\\ {\footnotesize{Institute
for Theoretical Physics, University of Bern,\\
Sidlerstrasse 5, CH-3012 Bern, Switzerland}} \\
\vspace*{0.5cm}
      and   \\
\vspace*{0.5cm}
      {\bf E. Pallante\footnotemark
      \footnotetext{ email: pallante@nbivax.nbi.dk}} \\
{\footnotesize{ NORDITA, Blegdamsvej 17,
2100 Copenhagen, Denmark}} \\   \end{center}

\vspace*{1.0cm}

\begin{abstract}

Within the Extended Nambu Jona-Lasinio model we analyse the
$1/N_c$-corrections to the leading order result $M_S=2M_Q$ where $M_Q$ is the
constituent quark mass.

\end{abstract} \vspace*{1.5cm} \begin{flushleft}
BUTP-95/7\\NORDITA 95/41 N,P  \\
May 1995 \end{flushleft} \end{titlepage} \newpage

Effective constituent quark models {\em \`a la} Nambu Jona-Lasinio
have recently been
used to a large extent to describe the low-energy  behaviour of
QCD and the hadron spectrum (see \cite{BIJNENS} for a  recent review).
Phenomenologically these models are found to be very successful in
reproducing the experimental values of the low-energy coupling constants
to $O(p^4)$ in Chiral Perturbation Theory ($\chi$PT)
\cite{ENJL}. Many of the couplings between resonances and pseudoscalar
mesons
have also been  computed \cite{ENJL,PRADES} and compare well with
experiment,  as well as the vector and axial-vector masses
\cite{ENJL,2point}.

When we turn our attention to the scalar sector the situation is less
satisfactory. Within the Extended Nambu Jona-Lasinio (ENJL) model the
scalar  two-point function shows up a pole at a mass $M_S = 2M_Q$
\cite{2point},  where $M_Q$ is the constituent quark
mass. With a typical value of $M_Q = 250\div 350$ MeV  (this range of
values is taken from the different fits of \cite{ENJL}) one has $M_S =
500\div 700$ MeV. Experimentally there is no evidence for so low a
scalar resonance. Until 1976 the Particle Data Book quoted a
scalar resonance  (called $\epsilon$) between 600 and 800 MeV with a
width of more than 600 MeV. Because of this
very large width it has then been discarded from the particle tables.
Unclear experimental signal of a narrow scalar state around 750 MeV
is reported in \cite{SVEC},
while the first  clear scalar resonances are the $a_0(983)$ and the
isosinglet $f_0(975)$ states. The proper $q\bar{q}$ assignment of the
$a_0(983)$ remains a problem. Its interpretation as ordinary $q\bar{q}$
system \cite{BRAMON} is not excluded; other possible interpretations are
 as a $q\bar{q}q\bar{q}$ state \cite{ACHASOV} or a $K\bar{K}$ bound
state \cite{WEINSTEIN}. For the $f_0(975)$ the analysis of ref.
\cite{ZOU} supports the prediction of the unitarized quark model which
interprets $f_0(975)$ as a $q\bar{q}$ resonance with a large admixture
of $K\bar{K}$ virtual state, while ref. \cite{MORGAN} almost excludes
its interpretation as a $K\bar{K}$ molecule and favours a conventional
Breit-Wigner-like structure. In both cases the interpretation of the
$f_0(975)$ as an ordinary $q\bar{q}$ state is almost excluded.
Recently, a fit of the available data has been performed \cite{TORN}
indicating that the $K\bar{K}$ component is large for both the $a_0(983)$
and $f_0(975)$ states.

In the large-$N_c$ limit of the ENJL model the isoscalar resonance
situated at $2M_Q$ is predicted to have a very large width of more
than 600 MeV \cite{HATSUDA}.
Assuming that such  a resonance exists,
qualitatively at least, there would be no surprise if
$1/N_c$-corrections  in relation to the large isoscalar width created
a non-degeneracy  between the isoscalar (the hypothetical $\epsilon$)
and the scalar octet (which could be the $a_0$).

This is the physical intuition we would like to precise. The framework
of the ENJL model offers a starting point to study the
$1/N_c$-corrections. At this point it is worth mentioning that the
NJL-like models contain most of the effective
low-energy models discussed in the literature as particular cases.
Therefore such an approach
lies in a quite general context.

The effective ENJL Lagrangian is given by \cite{ENJL}:

\beq\label{ENJL}
{\cal L}_{ENJL} ={\cal L}_{QCD}^{\Lambda_\chi} + {8\pi^2 G_S\over
N_c\Lambda_\chi^2} [(\bar{q}q)^2 - (\bar{q}\gamma_5q)^2]  -{8\pi^2
G_V\over N_c\Lambda_\chi^2} [(\bar{q}\gamma_\mu q)^2 -
(\bar{q}\gamma_\mu\gamma_5q)^2] . \eeq

\vskip 0.4cm

\noindent The problem of the connection between QCD and this  Lagrangian
has been addressed in \cite{ENJL}. The Lagrangian (\ref{ENJL}) is the
first term of a double expansion in $1/N_c$ and in  $1/\Lambda_\chi^2$,
$\Lambda_\chi\simeq 1$ GeV being the  ultraviolet cutoff of the low-energy
effective theory.

The effective action derived in \cite{ENJL} from the Lagrangian
(\ref{ENJL}) is subject to two types of $1/N_c$ corrections. The first
type will not be studied here. It
concerns the $1/N_c$ expansion mentioned above: at next-to-leading order in
$1/N_c$, there are other dimension six
fermionic operators compatible with the symmetries of the original QCD
Lagrangian. The second type appears when one considers loops made of
chains of quark bubbles, {\it i.e.} loops of mesons. In this article we
shall be interested in the latter.

In section \ref{NB} we explain the relation between  the quark-bubbles
resummation in the non-bosonized ENJL model and the bosonized version.
Then, in section \ref{2} we present the effective Lagrangian we shall use
in order to perform the loop calculation. The coupling constants of this
Lagrangian are derived within the ENJL model.
Section \ref{3} deals with the analysis of the $1/N_c$-corrections. We show the
corrections to the gap-equation, the mass-splitting  between the scalar
singlet and non-singlet, and the overall shift of the singlet.
Finally section \ref{CONC} is devoted to conclusions and remarks.

\section{\bf Bosonized versus non-bosonized version of the ENJL model.}

\lab{NB}

Let us consider the isoscalar two-point function
$\Pi(q^2)=i\int d^4x~ e^{iqx} \langle 0 \vert T S(x) S(0)\vert 0\rangle$
where $S(x)\equiv -{1\over \sqrt{2}}\bar q(x) q(x)$. This Green function
has been computed in the large-$N_c$ limit within the non-bosonized
version of the ENJL model using proper-time regularization with an
ultra-violet cut-off $\Lambda_\chi$. The result \cite{2point} is obtained
by resumming the geometrical series

\beq
\Pi(Q^2)^{(N_c\to\infty)} =\bar\Pi(Q^2)
\sum_{n=0}^{\infty}\left(g_S\bar\Pi(Q^2)\right)^n,
\eeq

\vskip 0.4cm

\noindent where

\beq
\dis \bar\Pi(Q^2)={1\over g_S} - (Q^2+(2M_Q)^2)Z_S(Q^2)
\eeq

\vskip 0.4cm

\noindent is the bare fermion loop diagram in the mean-field
approximation, $Z_S(Q^2)$ is the wave function renormalization constant

\beq
Z_S(Q^2)={N_c\over 16\pi^2}2 \int_0^1 ~d\alpha \Gamma \biggl ( 0,
{\alpha(1-\alpha)Q^2+M_Q^2\over\Lambda_\chi^2} \biggr )
\eeq

\vskip 0.4cm

\noindent with $\Gamma (0, \eps )=\int_\eps^\infty~ dz {1\over z} e^{-z} $
and $g_S={4\pi^2 G_S/ N_c \Lambda_\chi^2}$. One obtains

\beq \label{PiN}
\Pi(Q^2)^{(N_c\to\infty)}={\bar \Pi(Q^2)\over 1 - g_S \bar
\Pi(Q^2)}= -{1\over g_S}\left(1 - {(Z_S(Q^2) g_S)^{-1} \over
Q^2+(2M_Q)^2}\right).
\eeq

\vskip 0.4cm

The next-to-leading contribution in $1/N_c$ is built by
dressing the leading-$N_c$ scalar two-point function with one loop of
linear chains of constituent quark bubbles. We shall concentrate here on
the example of the self-energy diagram, as shown in Fig.1,
involving only the isoscalar coupling $g_S$.

These multiloop diagrams contain overlapping divergences. The
regularization procedure we use is as follows: first we
regularize the momenta running in the quark-bubbles and secondly
we regularize the remaining momentum which runs in the internal loop
made of quark-bubbles. This allows us to use the
expressions found in \cite{2point} for the two-point functions, like Eq.
(\ref{PiN}).  The divergences occurring in the loop calculation have to be
interpreted as physical divergences since the ENJL model is an effective
non-renormalizable theory in which all the momenta are cutoff at the
ultra-violet scale $\Lambda_\chi$.

Notice also that expressions like (\ref{PiN}) should not be used beyond
the convergence radius of the geometrical series, given in this case by
$q^2=-Q^2=(2M_Q)^2$. Since $2M_Q$ is of the order of $\Lambda_\chi$
it is legitimate to make the integration over the
internal loop, {\it i.e.} the bosonic loop, up to such energies.

In order to compute the self-energy diagram of Fig. 1 we need the
three-point function

\beq
\bar T(q,k)= \int d^4x~d^4y~ e^{iqx}e^{iqy} \langle 0 \vert T S(x)
S(y)S(0)\vert 0\rangle .
\eeq

\vskip 0.4cm

\noindent It is found to be

\beq
\bar T(q,k)= {N_c\over 16\pi^2} 4M_Q
\biggl [\Gamma\biggl ( 0,{M_Q^2\over\Lambda_\chi^2} \biggr ) -{2\over
3}\Gamma\biggl ( 1,{M_Q^2\over\Lambda_\chi^2} \biggr )\biggr ]
+O(k^2,q^2,(q-k)^2) .
\eeq

\vskip 0.4cm

\noindent We now proceed to the resummation $g_S \Pi =
{g_S \bar\Pi\over 1 - g_S \bar\Pi} + {1\over 1 - g_S \bar\Pi}g_s
\Sigma {1\over 1 - g_S \bar\Pi} + \cdots$ and get

\beq\label{Pi} \Pi(Q^2)=-{1\over g_S}\left(1 - {(Z_S(Q^2) g_S)^{-1}
\over  Q^2+(2M_Q)^2- \Sigma(Q^2)Z_S(Q^2)^{-1}}\right) ,\eeq

\vskip 0.4cm

\noindent with

\beq\label{Sigma} \Sigma(q^2)= {1\over 2}g_S^2 \int {d^4k\over (2\pi)^4}
{\bar T(q,k)}^2 {1 \over 1 - g_S \bar \Pi(k^2)}\, {1 \over 1 - g_S \bar
\Pi((q-k)^2)} . \eeq

\vskip 0.4cm

\noindent This expression has to be compared with the one obtained
within the bosonized approach, where we consider the three-scalars vertex
at lowest order in the derivative expansion with coupling $\lambda$:

\beq\label{Pibos} \Pi(Q^2)=-{1\over g_S}\left(1 - {(Z_S(0) g_S)^{-1}
\over  Q^2+(2M_Q)^2- {\lambda}^2\, I(Q^2)}\right) , \eeq

\vskip 0.4cm

\noindent where

\beq I(q^2)={1\over 2}\int {d^4k\over (2\pi)^4}  {1 \over k^2 -
(2M_Q)^2}\, {1 \over (q-k)^2 - (2M_Q)^2} . \eeq

\vskip 0.4cm

\noindent Considering now Eqs. (\ref{Sigma}) and (\ref{Pi})  and
substituting
$\bar T(q,k)$ with $\bar T(0,0)$ and $ Z_S(Q^2)$ with $Z_S(0)$, Eq. (\ref{Pi})
reduces to Eq. (\ref{Pibos}) with

\beq \lambda ={\bar T(0,0)\over Z_S(0)^{3/2}}. \eeq

\vskip 0.4cm \noindent This is nothing but the expression of the
three-scalars vertex coupling $\lambda_3/3!$ which is obtained in the
next section (Eq. \ref{COUP}),
within the effective action approach \footnote{notice however that $Z_S$
differs slightly from the expression given by the heat-kernel calculation;
see \cite{2point} about this question.}.
Numerically the error made in approximating $T(q,k)$ with $T(q,0)$ is
of the order of thirty percent when $k^2$ is below $\Lambda_\chi^2$.
As a consequence of this approximation we shall not identify
the cut-off used in the regularization of the internal loop,
{\it i.e.} the bosonic loop, with $\Lambda_\chi$,
but rather keep it
as an adjustable parameter whose order of magnitude is
$\Lambda \sim 2M_Q \sim \Lambda_\chi$.

In what follows
we shall use the bosonized ENJL model where, for each vertex, only the
lowest order in the derivative expansion will be kept.
This approximation allows us to simplify the calculations and preserves
chiral invariance.

\section{\bf The effective Lagrangian}

\lab{2}

We restrict ourselves to the $SU(2)_L\times SU(2)_R$ case.
The general form for the meson matrices, singlets or triplets under
$SU(2)_V$, reads

\beq M= \sum_{a=1}^8 {1\over \sqrt{2}} M_{(a)}\tau^{(a)} + {1\over
\sqrt{2}}M_0 {\bf 1}, \eeq

\noindent where $\tau^{(a)}, a=1,..3$ are the Pauli matrices with $Tr
(\tau^a\tau^b) = 2\delta^{ab}$ and $M_0$ is the singlet component.
We use for the pseudoscalar field the exponential parametrization
$\xi(\Phi) = \exp (-{i\over \sqrt{2}}{\Phi\over f_\pi})$.

In the chiral limit ($m_u=m_d=0$), the effective chiral Lagrangian
including scalar and pseudoscalar mesons to $O(p^2)$ is given by:

\br
{\cal{L}}^{S,P}={f_\pi^2\over 4} <\xi_\mu \xi^\mu>
+{1\over 2}<d_\mu S d^\mu S>
-{1\over 2}M_S^2<S^2>+{\cal{L}}_{int}^{S,P} \\ \dis
{\cal{L}}_{int}^{S,P} =  -{\lambda_3\over 3!}<S^3> - {\lambda_4\over
4!}<S^4>   + c_d <S\xi_\mu\xi^\mu > + c_4^{(1)} <S^2\xi_\mu\xi^\mu >
+ c_4^{(2)} <S\xi_\mu S\xi^\mu > ,
\er

\noindent where $S$ is the scalar meson field and $\xi_\mu = i\{\xi^\dagger
(\partial_\mu -ir_\mu )\xi - \xi (\partial_\mu -il_\mu )\xi^\dagger\}$
 is the axial current constructed with the pseudoscalar field, where
$r_\mu$ and $l_\mu$ are the external right-handed and left-handed sources.
The couplings among mesons are
obtained by integrating out the constituent quark fields in the
bosonized ENJL model using the heat kernel
expansion with proper time regularization  as explained in ref.
\cite{ENJL}. They are functions of the ultra-violet cut-off $\Lambda_\chi$,
the constituent quark mass $M_Q$, the axial-pseudoscalar mixing parameter
$g_A$ and  the number of colours $N_c$. We find:

\begin{eqnarray} {\lambda_3\over 3!} &=&{N_c\over 16\pi^2}4{M_Q\over
Z_S^{3/2}} \biggl [\Gamma (0,\eps ) -{2\over 3}\Gamma (1,\eps )\biggr ]
\nonumber\\ {\lambda_4\over 4!} &=&{N_c\over 16\pi^2}{1\over Z_S^2}
\biggl [\Gamma (0,\eps )-4\Gamma (1,\eps ) +{4\over 3} \Gamma (2,\eps
)\biggr ] \nonumber\\ c_d&=&  {N_c\over 16\pi^2} M_Q{2g_A^2\over
\sqrt{Z_S}} \biggl [\Gamma (0,\eps )-\Gamma (1,\eps ) \biggr ]
\nonumber\\ c_4^{(1)}&=& {1\over 2} {N_c\over 16\pi^2} {g_A^2\over Z_S}
\biggl [\Gamma (0,\eps ) -{20\over 3}  \Gamma (1,\eps ) +{8\over 3}
\Gamma (2,\eps )\biggr ] \nonumber\\ c_4^{(2)}&=&{1\over 2} {N_c\over
16\pi^2} {g_A^2\over Z_S} \biggl [\Gamma (0,\eps ) -{10\over 3}  \Gamma
(1,\eps ) +{4\over 3} \Gamma (2,\eps )\biggr ] .
\lab{COUP}
\end{eqnarray}

\noindent The incomplete gamma functions $\Gamma (n-2, \eps )$, with
$\eps = M_Q^2/\Lambda_\chi^2$, are defined as

\beq \Gamma (n-2, \eps )=\int_\eps^\infty~ dz {1\over z} e^{-z} z^{n-2}.
\lab{GAMMA} \eeq

\noindent $\Gamma (-2, \eps )$ contains a quartic divergence, $\Gamma
(-1, \eps )$ a quadratic one, $\Gamma (0, \eps )$ is logarithmically
divergent,  while $\Gamma (n, \eps )$ with $n>0$ are finite.

The Interaction Lagrangian of
the scalar mesons with vectors and  axial-vectors at the first two leading
chiral orders, {\it i.e.} $O(p^0)$ and $O(p^2)$, is\footnote{
The Lagrangian involving vectors starts at $O(p^2)$. For consistency we need
to include $O(p^0)$ as well as $O(p^2)$ terms in the Lagrangian involving
axial-vectors.}:

\begin{eqnarray} {\cal{L}}_{int}^{V,A} &=&
 \tilde{c}_A <SA_\mu A^{\mu} > + c_A^{(1)}
<SA_\mu SA^{\mu} >+ c_A^{(2)} <S^2A_\mu A^{\mu} >
\nonumber\\
&&+ c_{AP} <S\{\xi_\mu ,A^\mu\} >
+  c_V^{(1)}<SV_\mu SV^{\mu} >  + c_V^{(2)} <S^2V_\mu V^{\mu}>
\nonumber\\ &&+ \tilde d_V <S
V_{\mu\nu}  V^{\mu\nu}> + d_V^{(1)} <S V_{\mu\nu} SV^{\mu\nu}>
+ d_V^{(2)} <S^2 V_{\mu\nu} V^{\mu\nu}>
\nonumber\\ &&+ \tilde d_A <S A_{\mu\nu}  A^{\mu\nu}>
+ d_A^{(1)} <S A_{\mu\nu} SA^{\mu\nu}>
+ d_A^{(2)} <S^2 A_{\mu\nu} A^{\mu\nu}>.
\end{eqnarray}

\noindent Notice also the presence at $O(p^0)$ of the mixed term
scalar-pseudoscalar-axial with coupling $c_{AP}$.
$Z_V$ and $Z_A$ are the wave function renormalization constants
of the vector and axial-vector fields:

\beq Z_V = Z_A = {N_c\over 16\pi^2}{1\over 3}\Gamma (0,\eps ) \eeq

\noindent The couplings are found to be:

\begin{eqnarray} \tilde{c}_A&=& {1\over 2} {N_c\over 16\pi^2}
{M_Q\over  \sqrt{Z_S}Z_V}
 \biggl [4\Gamma (0,\eps ) -4\Gamma (1,\eps ) \biggr ] \nonumber\\
c_A^{(1)}&=&{1\over 2} {N_c\over 16\pi^2}{1\over Z_S Z_V}
 \biggl [\Gamma (0,\eps ) -{10\over 3} \Gamma (1,\eps ) +{4\over 3}
\Gamma (2,\eps )\biggr ] \nonumber\\
c_A^{(2)}&=&{1\over 2} {N_c\over
16\pi^2}{1\over Z_S Z_V}
 \biggl [\Gamma (0,\eps ) -{20\over 3} \Gamma (1,\eps ) +{8\over 3}
\Gamma (2,\eps )\biggr ] \nonumber\\
c_{AP}&=&{1\over 2} {N_c\over 16\pi^2}{g_A M_Q\over \sqrt{Z_S}\sqrt{
Z_V}}
 \biggl [ -4\Gamma (0,\eps ) +4 \Gamma (1,\eps ) \biggr ]
\nonumber\\
c_V^{(1)}&=&-c_V^{(2)} ={1\over 2} {N_c\over
16\pi^2}{1\over Z_S Z_V}  \biggl [-\Gamma (0,\eps )+{2\over 3} \Gamma
(1,\eps ) \biggr ] \nonumber\\
\tilde d_V&=&{1\over 6} {N_c\over 16\pi^2}{1\over M_Q}{1\over
\sqrt{Z_S} Z_V}
 \Gamma (1,\eps )  \nonumber\\
\tilde d_A&=&{N_c\over 16\pi^2}{1\over M_Q}{1\over
\sqrt{Z_S} Z_V}
 \biggl [{1\over 2} \Gamma (1,\eps )-{1\over 3} \Gamma (2,\eps )\biggr ]
  \nonumber\\
d_V^{(1)}&=&-{1\over 2} {N_c\over 16\pi^2}{1\over
Z_S Z_V}{1\over M_Q^2} \biggl [{1\over 6}\Gamma (1,\eps )+
{2\over 15} \Gamma (2,\eps ) \biggr ] \nonumber\\
d_A^{(1)}&=&-{1\over 2} {N_c\over 16\pi^2}{1\over
Z_S Z_V}{1\over M_Q^2}
\biggl [ -{1\over 6} \Gamma (1,\eps ) +{7\over 15} \Gamma (2,\eps
)\biggr ] \nonumber\\
d_V^{(2)}&=&-{1\over 2} {N_c\over
16\pi^2}{1\over Z_S Z_V}{1\over M_Q^2}
 \biggl [ -{1\over 3} \Gamma (1,\eps ) +{1\over 5} \Gamma (2,\eps
)\biggr ] \nonumber\\
d_A^{(2)}&=&-{1\over 2} {N_c\over
16\pi^2}{1\over Z_S Z_V}{1\over M_Q^2}
 \biggl [ -{1\over 3} \Gamma (1,\eps ) +{23\over 15} \Gamma (2,\eps
)\biggr ] .
\end{eqnarray}

\section{\bf Analysis of the $1/N_c$ corrections}

\lab{3}

In the Appendix we have listed the results
for all the loop contributions. We
work in the chiral limit  ($m_\pi =0$) and in the zero momentum limit
($q^2=0$). The masses of the vector and axial-vector mesons have been
fixed to  their experimental value: $M_V\simeq 0.770$ GeV \cite{3point} and
$M_A\simeq 1.2$ GeV (we disregard the error on the axial mass).

We have calculated the $1/N_c$ corrections in two cases:
1) assuming that the
scalar particle is a singlet (which, as we said in the introduction, cannot be
identified with the $f_0(975)$ resonance) 2) assuming that the quark
content of the scalar particle is the same as that of the $\rho (770)$
vector meson. This could be the case of the physical $a_0 (983)$ scalar
resonance. Obviously our $SU(2)$ calculation has to be
interpreted as  a first indicative approximation of the fully realistic
$SU(3)$  calculation.

The diagrams which contribute to the scalar two-point function at
next-to-leading order in $1/N_c$ are shown in Fig. 2. They are the
self-energy (a), the tadpole (b) and the `top' (c) diagrams.

Particles running in the loop are the singlet and the $SU(2)$ triplet
states for each quantum number: Scalar, Pseudoscalar, Vector, and
Axial-Vector. We disregard in the present analysis the splitting
between the singlet and the triplet pseudoscalar states; this is due
to the $U(1)$ axial anomaly, which appears in the effective Lagrangian
at next-to-leading order in the $1/N_c$ expansion.

The correction to the mass depends quartically on the UV cut-off $\Lambda$
of the bosonic loop. It is proportional to

\beq
{1\over N_c}\left(\Lambda\over 2M_Q\right)^4.
\eeq

\vskip 0.4cm

\noindent Hence reasonable values of $\Lambda$ cannot exceed $2M_Q$.

\vskip 0.4cm

Our conclusions are threefold:

1) The {\em top} diagram can be alternatively included in the corrections to
the scalar propagator or in the gap-equation so that they translate into
corrections to the leading-$N_c$ value of $M_Q$.
These corrections do not
create any mass splitting between singlet and non-singlet states.
In addition
they are $q^2$ independent. The main result concerning these diagrams
is the compensation between negative contributions (pseudoscalar and scalar)
and positive contributions (vector and axial)
so that the corrections to the gap-equation are small.

In Ref. \cite{Ebert} an attempt to compute the $1/N_c$-corrections
to the gap equation of the NJL model has been
performed in the $G_V=0$ case. However, the formalism they have adopted
implies non-derivative couplings of pseudoscalar mesons to other particles.
This makes the comparison with our approach problematic.
In particular, the relative size and sign of
the scalar and pseudoscalar contributions cannot be compared in the two cases.


2) The mass splitting can arise from the {\em self-energy} diagrams and
the {\em tadpole} diagrams, but not from the {\em top} diagrams.
Our conclusion  is that the $1/N_c$-corrections make the non-singlet
heavier than the singlet.

3) Both in the scalar singlet and non singlet cases the
corrections are negative and quite sensitive to the value of the
UV cut-off. Positive values of the renormalized scalar singlet  mass
are limited to low values of $\Lambda$.
We have considered the following ranges for the different parameters:

\br
250\,\hbox{MeV} \le M_Q \le 350\,\hbox{MeV}\\\dis
0.6\le g_A\le 0.8\\\dis
900\,\hbox{MeV}\le \Lambda_\chi\le 1100\,\hbox{MeV}\\\dis
500\,\hbox{MeV}\le \Lambda\le 2M_Q .
\er

\vskip 0.4cm

\noindent For $M_Q\sim 250$ MeV
corrections are big in absolute value (more than 100$\%$).
For $M_Q\sim 350$ MeV
and $\Lambda\sim 500$ MeV
we obtain smaller  corrections but the mass splitting is also smaller.
With $M_Q=350$ MeV, $\Lambda_\chi=900$ MeV, $g_A=0.7$ and
$\Lambda=500$ MeV we find for the singlet $M_{s}\simeq 500$ MeV and
for the non-singlet $M_{ns}\simeq 560$ MeV.

\section {\bf Conclusions} \lab{CONC}

Our work is a first attempt to obtain an order of magnitude as well as the sign
of the $1/N_c$ corrections to the mass of the
scalar resonance in the framework of the ENJL model. A part from the problem
of overlapping divergences, there is a definite way of computing these
corrections in the full non-bosonized ENJL model.
As explained in Section 1 we have performed the calculations within
the bosonized ENJL version.

Our main result is the existence at next-to-leading order in $1/N_c$ of a
mass-splitting between singlet and non-singlet states,
making the non-singlet heavier.

We find that results are extremely sensitive to the values
of the UV cut-off $\Lambda$ of the bosonic loop.
Corrections to the singlet mass are negative.
However, contrary to the
mass-splitting, the overall shift of the singlet can be affected by
corrections coming from higher dimensional fermionic operators. Whereas these
corrections are about 30$\%$ in the vector sector \cite{QR} their
importance for the scalar sector is not known.

\vspace{3.cm}
{\bf{Acknowledgements}} \vspace{0.8cm}

It is a pleasure to thank J. Bijnens for having called our attention
to this problem, for many useful discussions and a careful reading of
the manuscript. We also thank J. Gasser, H. Leutwyler,
P. Minkowski and R. Petronzio for interesting and stimulating discussions.

The work by CB was supported by the EU Contract Nr. ERBCHRXCT 920026.
The work by EP was supported by the EU Contract Nr. ERBCHBGCT 930442.
\newpage

\appendix

{\large \bf  Appendix}

\vskip 1cm

{\large\bf Top diagrams}

\begin{eqnarray}
S &&i{\lambda_3^2\over 4}{1+3\over 16\pi ^2}\Gamma ( -1, M_S)\nonumber\\
P && i{\lambda_3 c_d\over 2}{2\over f_\pi^2} {1\over M_S^2} {1+3\over
16\pi ^2}2 (m_\pi^2)^2\Gamma ( -2, m_\pi)\nonumber\\
A\, O(p^0)&& i{\lambda_3 \tilde{c}_A\over 2}{1+3\over
16\pi ^2} {M_A^2\over M_S^2} \biggl [2 \Gamma(-2,M_A)+ 4 \Gamma
(-1,M_A)\biggr ]\nonumber\\
A(V) \, O(p^2)&& -i\lambda_3\tilde d_{A(V)} 3{1\over M_S^2}
{1+3\over 16\pi ^2}2 M_{A(V)}^4\Gamma (-2,M_{A(V)})
\end{eqnarray}

\vskip 0.4cm

\noindent Contributions are the same for singlet and triplet propagator.

\vskip 1cm

{\large\bf Self-energy diagrams (at $q^2=0$)}

\vskip 0.4cm

{\bf Singlet propagator}

\begin{eqnarray}
S && i{\lambda_3^2\over 4} {1+3\over 16\pi ^2}
\Gamma ( 0, M_S) \nonumber\\
P && ic_d^2 \biggl ({2\over f_\pi^2}\biggr )^2{1+3\over 16\pi ^2} \biggl
[2m_\pi^4 \Gamma ( -2,m_\pi)
-m_\pi^4 \Gamma( -1,m_\pi )+m_\pi^4 \Gamma ( 0,m_\pi)\biggr]\nonumber\\
A\, O(p^0)\times O(p^0) && i(\tilde{c}_A)^2 {1+3\over 16\pi ^2} \biggl
[2\Gamma ( -2,M_A) + \Gamma ( -1, M_A) +3\Gamma  (0, M_A)\biggr ]
\nonumber\\
A\, O(p^0)\times O(p^2) &&- 2i\tilde c_A \tilde d_A {1+3\over 16\pi ^2}
6M_A^2\biggl[ \Gamma ( -1, M_A) - \Gamma (0, M_A)\biggr ]
\nonumber\\
A(V)\, O(p^2)\times O(p^2) && i\tilde d_{A(V)}^2 {1+3\over 16\pi ^2}
12M_{A(V)}^4\biggl[ 2\Gamma (-2,M_{A(V)})- \Gamma
(-1,M_{A(V)}) + \Gamma(0,M_{A(V)})\biggr]\nonumber\\
A-P && 2i c_{AP}^2{2\over f_\pi^2}{1+3\over 16\pi ^2}{1\over M_A^2}
2m_\pi^4 \Gamma ( -2,m_\pi) \nonumber\\
\end{eqnarray}

\vskip 0.4cm

{\bf Triplet propagator}

\vskip 0.4cm

\noindent  The only possible self-energy diagrams for the triplet
propagator contain two different internal lines, one singlet and one
triplet. The
contribution is half the contribution to the singlet propagator displayed
above.

\vskip 1cm

{\large\bf Tadpole diagrams}

\vskip 0.4cm

{\bf Singlet propagator}

\begin{eqnarray}
S &&-i{\lambda_4\over 4} M_S^2 {1+3\over 16\pi ^2}\Gamma ( -1, M_S),
\nonumber\\
P && -i(c_4^{(1)}+c_4^{(2)}){2\over f_\pi^2}
{1+3\over 16\pi ^2}2 m_\pi^4\Gamma ( -2, m_\pi) \nonumber\\
A(V)\, O(p^0)&& -i(c_{A(V)}^{(1)}+c_{A(V)}^{(2)})
{1+3\over 16\pi ^2}  M_{A(V)}^2
\biggl [ 2\Gamma (-2,M_{A(V)}) +4 \Gamma (-1,M_{A(V)})\biggr ]\nonumber\\
A(V) \, O(p^2)&&i(d_{A(V)}^{(1)}+d_{A(V)}^{(2)})6{1+3\over 16\pi ^2}
2 M_{A(V)}^4\Gamma (-2,M_{A(V)})
\end{eqnarray}

\vskip 0.4cm

\noindent Notice that in the singlet case there are no contributions  from
vector vertices of order $O(p^0)$ like $<SV_\mu V^{\mu}>$, because
$c_V^{(1)}+c_V^{(2)} =0$.

\vskip 0.4cm

{\bf Triplet propagator}

\vskip 0.4cm

\begin{eqnarray} S &&-i\lambda_4({1\over 4}+{1\over 4}+{1\over 6})
M_S^2{1\over 16\pi ^2} \Gamma
( -1, M_S) \nonumber\\
P &&-i{2\over f_\pi^2}( 3 c_4^{(1)}- c_4^{(2)}+ c_4^{(1)}+
c_4^{(2)} ){1\over 16\pi ^2}2 m_\pi^4\Gamma ( -2,
m_\pi) \nonumber\\
A(V) \,O(p^0) &&-i(-c_{A(V)}^{(1)}+3c_{A(V)}^{(2)}
+ c_{A(V)}^{(1)}+ c_{A(V)}^{(2)} )
{M_{A(V)}^2\over 16\pi^2} \biggl [ 2 \Gamma (-2,M_{A(V)})+4  \Gamma
(-1,M_{A(V)})\biggr ] \nonumber\\
A(V)\,O(p^2) && i(3 d_{A(V)}^{(2)} - d_{A(V)}^{(1)} + d_{A(V)}^{(2)}
+ d_{A(V)}^{(1)} ) 6 {1\over 16\pi ^2}2 M_{A(V)}^4\Gamma (-2,M_{A(V)})
\end{eqnarray}

\vfill\eject

\centerline {{\bf FIGURE CAPTIONS}} \vskip2.truecm

\begin{description}

\item[1)] Self-energy diagram built by
dressing the leading-$N_c$ scalar two-point function with one loop of
linear chains of constituent quark bubbles.

\item[2)] One loop diagrams which give the next-to-leading in $1/N_c$
contribution to the scalar propagator: (a) the self-energy diagram, (b)
the tadpole diagram and (c) the ``top'' diagram. The solid line is a
scalar field, while the dashed line in the loop can be any mesonic field.

\end{description}

\vfill\eject

\begin{figure}
\unitlength 1cm
\begin{picture}(7,5)(-8,-2.5)
\thicklines
\multiput(-4,0)(0.6,0){4}{\circle{0.6}}
\multiput(-2.2,-1.2)(0,0.6){5}{\circle{0.6}}
\multiput(2.0,-1.2)(0,0.6){5}{\circle{0.6}}
\multiput(-2.2,-1.2)(0.6,0){8}{\circle{0.6}}
\multiput(-2.2,1.2)(0.6,0){8}{\circle{0.6}}
\multiput(2.0,0)(0.6,0){4}{\circle{0.6}}

\put(-0.5,-4){}
\end{picture}
\caption{}
\end{figure}

\begin{figure}
\epsfig{file=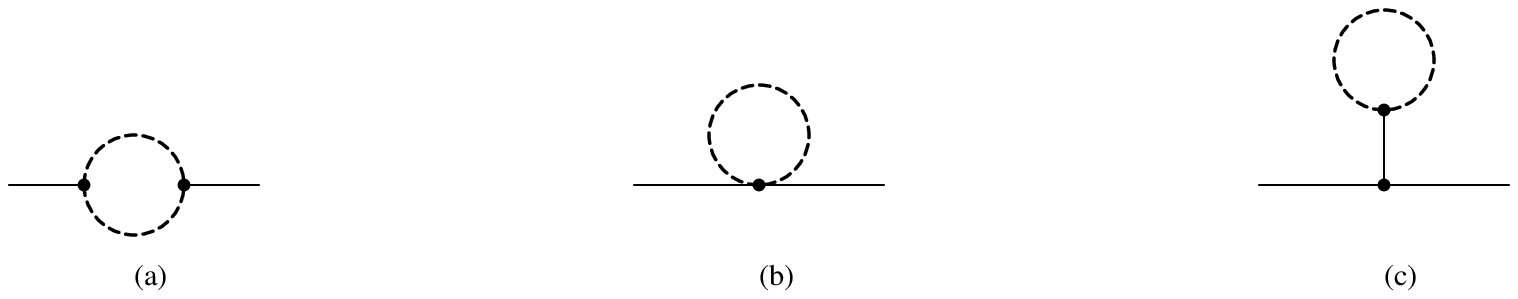}
\caption{}
\end{figure}

\end{document}